\pdfoutput=1
\documentclass[english,aps,prl,twocolumn, superscriptaddress,showpacs, amsmath]{revtex4-1}

\usepackage[T1]{fontenc}
\usepackage[latin9]{inputenc}
\usepackage{color}
\usepackage{units}
\usepackage{amssymb}
\usepackage{graphicx}
\usepackage{esint}
\usepackage{bm}
\usepackage{natbib}
\usepackage{ulem}
\usepackage[colorlinks=true, citecolor=red]{hyperref}
\setcitestyle{journalcolor= blue}

\newcommand{\vl}{$v^l$ }
\newcommand{\vnl}{$v^{nl}$ }
\newcommand{\noise}{$\langle\delta R^2\rangle/ \langle R\rangle^2$ }

\newcommand{\PSD}{power spectral density}
\usepackage{graphicx}
\usepackage{dcolumn}
\usepackage{bm}

\usepackage{babel}

\begin{document}

\title {Probing the interplay between surface and bulk states in the topological Kondo insulator SmB$_6$ through conductance fluctuation spectroscopy}
\author{Sangram Biswas}

\affiliation{Department of Physics, Indian Institute of Science, Bangalore 560012, India}

\author{M. Ciomaga Hatnean}
\affiliation{Department of Physics, University of Warwick, Coventry, CV4 7AL, UK.}

\author{G. Balakrishnan}
\affiliation{Department of Physics, University of Warwick, Coventry, CV4 7AL, UK.}

\author{Aveek Bid}
\email{aveek.bid@physics.iisc.ernet.in}
\affiliation{Department of Physics, Indian Institute of Science, Bangalore 560012, India}

\begin{abstract}

We present results of resistance fluctuation spectroscopy on single crystals of the predicted Kondo topological insulator material SmB$_6$. Our measurements show that at low temperatures, transport in this system takes place only through surface states. The measured noise in this temperature range arises due to Universal Conductance Fluctuations whose statistics was found to be consistent with theoretical predictions for that of two-dimensional systems in the Symplectic symmetry class. At higher temperatures, we find signatures of glassy dynamics and establish that the measured noise is caused by mobility fluctuations in the bulk. We find that, unlike the topological insulators of the dichalcogenide family, the noise in surface and bulk conduction channels in SmB$_6$ are completely uncorrelated. Our measurements establish that at sufficiently low temperatures, the bulk has no discernible contribution to electrical transport in  SmB$_6$ making it an ideal platform for probing the physics of topological surface states. 
 
\end{abstract}

\maketitle

\section{Introduction}
The rare-earth hexaboride SmB$_6$, a strongly correlated heavy fermion Kondo insulator, has shot to recent prominence because of the prediction that it can support topologically protected surface states. This prediction is based on the following line of reasoning:  at temperatures below the Kondo energy scale $T_K$, a narrow gap ($\sim$ 3-5 meV) opens up at the Fermi energy because of the hybridization of the localized Sm $4f$ band with the dispersive Sm $5d$ conduction band~\cite{PhysRevLett.104.106408, PhysRevLett.110.096401, PhysRevLett.112.136401, kim2013surface, kong2011tuning,wolgast2013low, neupane2013surface, jiang2013observation, xu2014direct, xu2014exotic}. Microscopically, this happens because of the screening of the individual Sm$^{3+}$ 4$f^5$ local moments by the itinerant 5$d$ electrons. 

The process of formation of this Kondo gap involves the transfer of electrons, at temperatures below $T_K$, from Sm $4f$ band to Sm $5d$ band which causes SmB$_6$ to be a `mixed valence' compound~\cite{PhysRevB.66.165209}. Consequent to this charge transfer, there is a possibility of band inversion between $4f$ and $5d$ orbitals. The fact that this happens an odd number of times in SmB$_6$~\cite{PhysRevLett.110.096401, park2016topological} has lead to the intriguing possibility that this material may be the first realization of a topological insulator in a strongly interacting system~\cite{PhysRevLett.110.096401}. The existence of surface states (SS) in SmB$_6$ at temperatures below $\sim$~5~K is supported by several experiments including angle-resolved photoemission spectroscopy~\cite{neupane2013surface,xu2014exotic,xu2014direct} (ARPES),  electrical transport~\cite{syers2015tuning,kim2014topological}, Hall measurements~\cite{kim2013surface}, point contact spectroscopy~\cite{flachbart2001energy,zhang2013hybridization} and cantilever magnetometry~\cite{li2014two} although there is no unambiguous proof of their topological origin. 

In this article we present results of high resolution resistance fluctuation (noise) spectroscopy in single crystal samples of SmB$_6$ in both local and non-local configurations. The primary aim of these  experiments was to probe the nature of charge scattering in the surface states in SmB$_6$ and to understand their effect on electrical transport  as the system undergoes a transition from a Kondo insulator(KI) to a spin polarized Dirac metal. 
We find that at ultra-low temperatures the noise is dominated by Universal Conductance Fluctuations (UCF) with a 2-dimensional nature. At higher temperatures, the noise comes from local moment fluctuations in the Kondo cloud. Our study establishes that noise measurements are much more effective than traditional transport measurements in detecting signatures of surface transport in this class of systems.

The single crystal samples used in these experiments were grown by floating zone technique using a high-power xenon arc lamp image furnace~\cite{hatnean2013large}. The measurements were carried out on two different crystal pieces cut from the same crystal boule - the results obtained from both the samples were quantitatively similar. Electrical contacts separated by 200~$\mu m$ were defined on the (110) surface by deposition of Cr/Au pads. Before defining the contacts, the (110) surface was mirror-polished and cleaned using concentrated HCl to get rid of surface contaminants. 

\begin{figure}[h!]
	\begin{center}
		\includegraphics[width=0.45\textwidth]{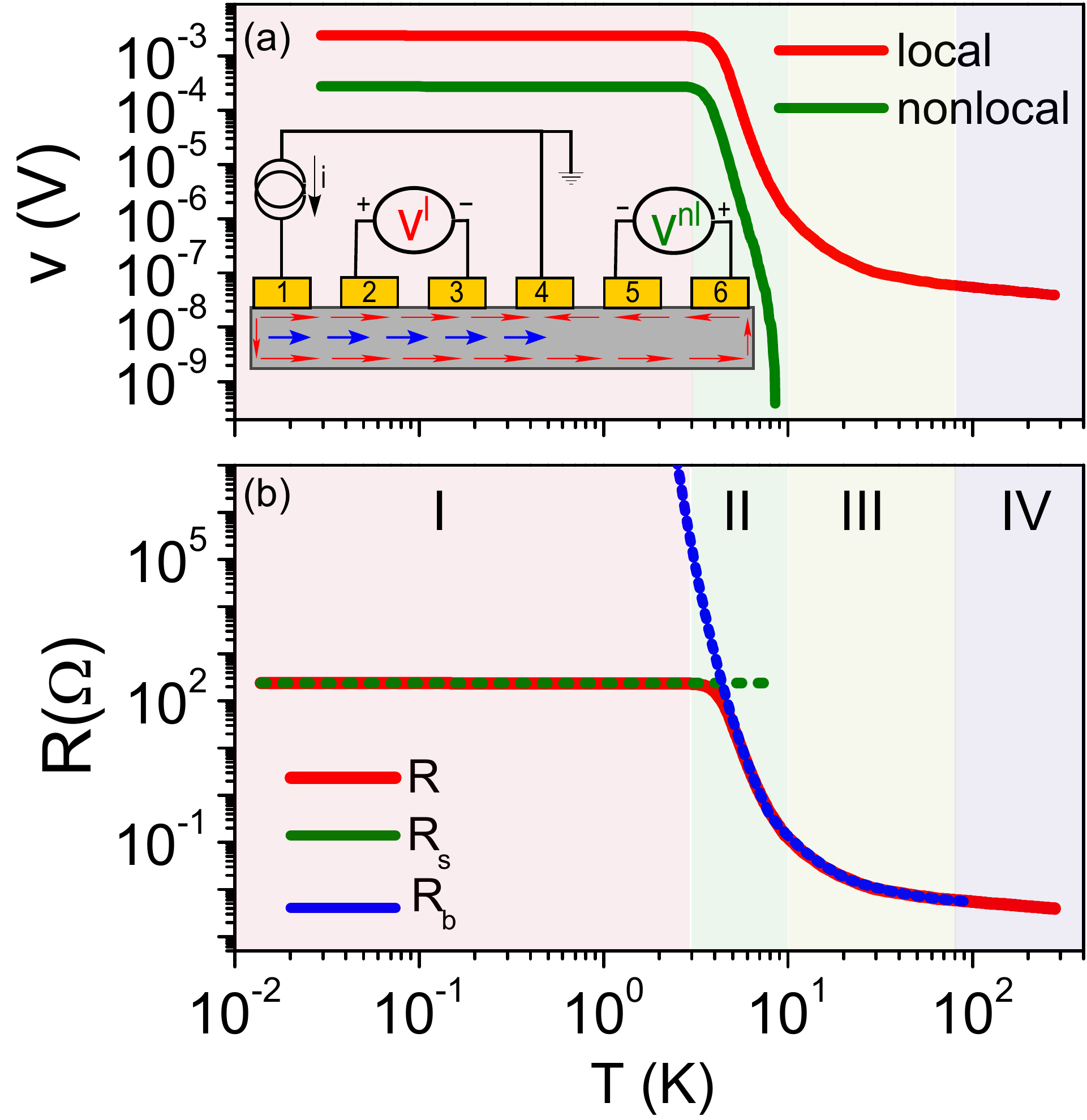}
		\small{\caption{ (color online) (a) The main panel shows a plot of the measured local and non-local voltages as a function of temperature. The inset shows a schematic of device contact configuration. (b) Plots of $R_s$ (olive dotted line) and $R_b$ (blue dotted line) as  a function of temperature. The solid red line shows the measured $v^l/i$. In both the panels, the shading indicates the four different temperature regimes as explained in detail in the text. 
				\label{fig:RT}}}
	\end{center}
\end{figure}

The inset in figure~\ref{fig:RT}(a) shows a schematic of the device and defines the local and non-local transport configurations. The directions of bulk and surface currents are shown by blue and red arrows respectively. One part of the surface current flows from the source (contact~\#1) to the drain (contact~\#4) through the top surface via contacts 2 and 3. This part of the surface current, combined with the bulk current, generates the local voltage \vl between the contacts 2 and 3. Another part of the surface current, as shown by the schematic, flows successively through the left side surface, bottom surface, right side surface, and back to the top surface before being collected at the drain (contact~\#4). It is this part of surface current which gives us finite non-local voltage \vnl between contacts 6 and 5.

Figure~\ref{fig:RT}(a) shows a plot of both the local \vl and the non-local  \vnl voltages as a function of temperature. The measurements were performed by applying a $10~\mu A$ ac current through the sample using standard 4-probe configuration.  The local voltage \vl increased by more than four orders of magnitude as the sample was cooled down from room temperature which attests to the high purity of the SmB$_6$ crystals. The non-local voltage appeared below $\sim 10~K$ and increased rapidly by more than six orders of magnitude before saturating below $T=3~K$. We have shown in a previous publication that the appearance of this non-local transport signal can be attributed to SS which creates an additional transport channel in parallel with the bulk of the sample~\cite{biswas2015robust}. Using the method described in Ref.~\cite{biswas2015robust}, the values of $R_s$ and $R_b$ were  extracted from the measured \vl and are plotted in figure~\ref{fig:RT}(b). 
\begin{figure}[t!]
	\begin{center}
		\includegraphics[width=0.45\textwidth]{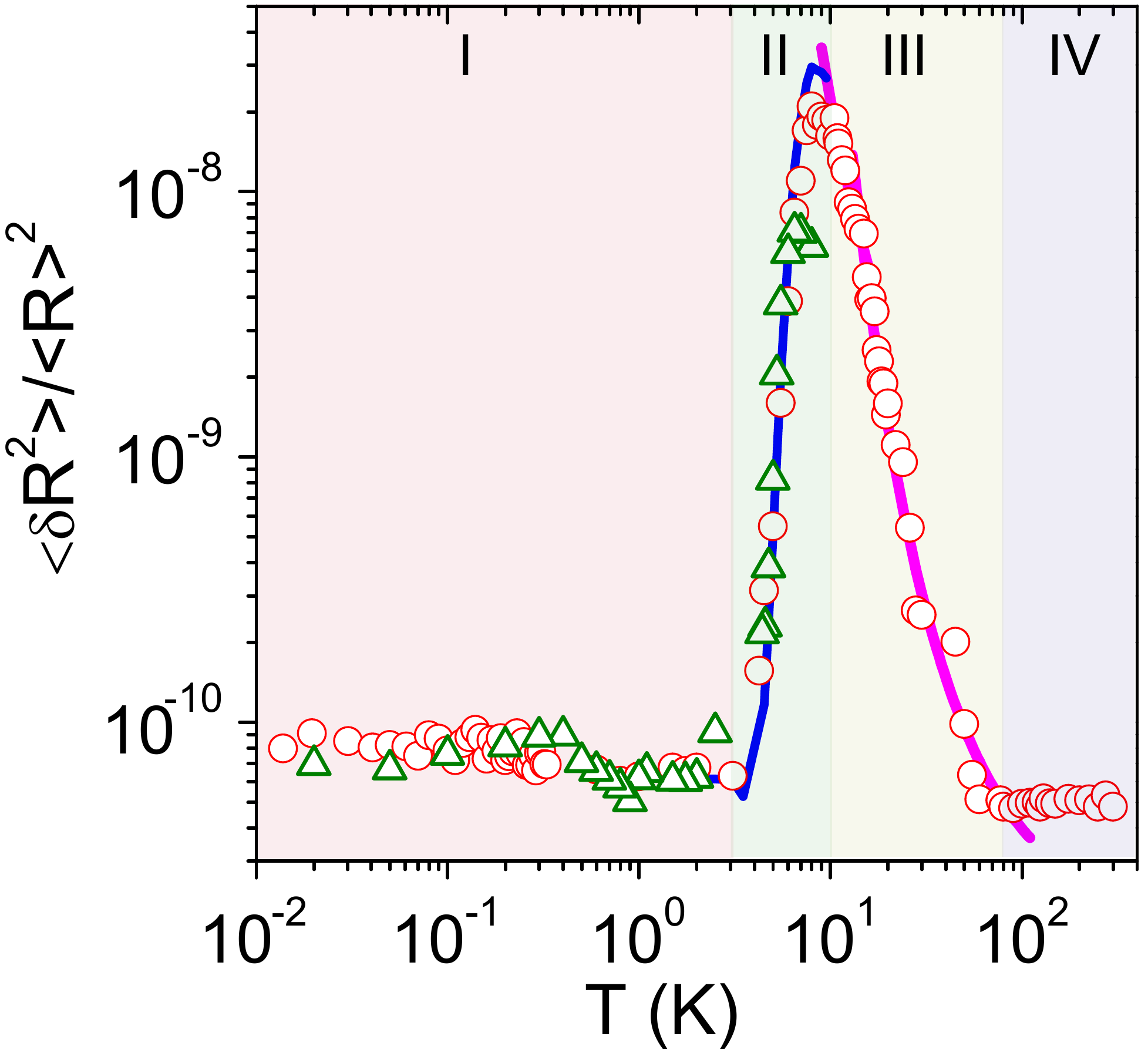}
		\small{\caption{(color online) (Local (red open circles) and non-local (olive open triangles) resistance fluctuations \noise as a function of temperature for sample S2. The blue line shows the fit to Eqn.~\ref{eqn:noisediv}. The pink line is a fit to the form \noise $\propto 1/n^{0.8}$.   
				\label{fig:noise}}}
	\end{center}
\end{figure}
To further probe the dynamics of SS and bulk conduction channels in this system, we performed low frequency noise measurements over the temperature range 10~mK - 300~K  using a digital signal processing (DSP) based ac 4-probe technique (Details of the measurement technique are provided in Appendix~A). This technique allows a simultaneous measurement of the Johnson-Nyquist background noise as well as the bias dependent noise from the sample~\cite{ghosh2004set,scofield1987ac}. At every temperature, the time series of resistance fluctuations was accumulated using a fast 16-bit analog-to-digital conversion card from which the power spectral density (PSD) of voltage fluctuations, $S_V(f)$ was calculated. The PSD was integrated over the band-width of measurement to obtain the relative variance of resistance fluctuations \noise at a given temperature, \noise$=\int{S_V(f)df}/\langle V\rangle^2$.

\begin{figure}[t!]
	\begin{center}
		\includegraphics[width=0.45\textwidth]{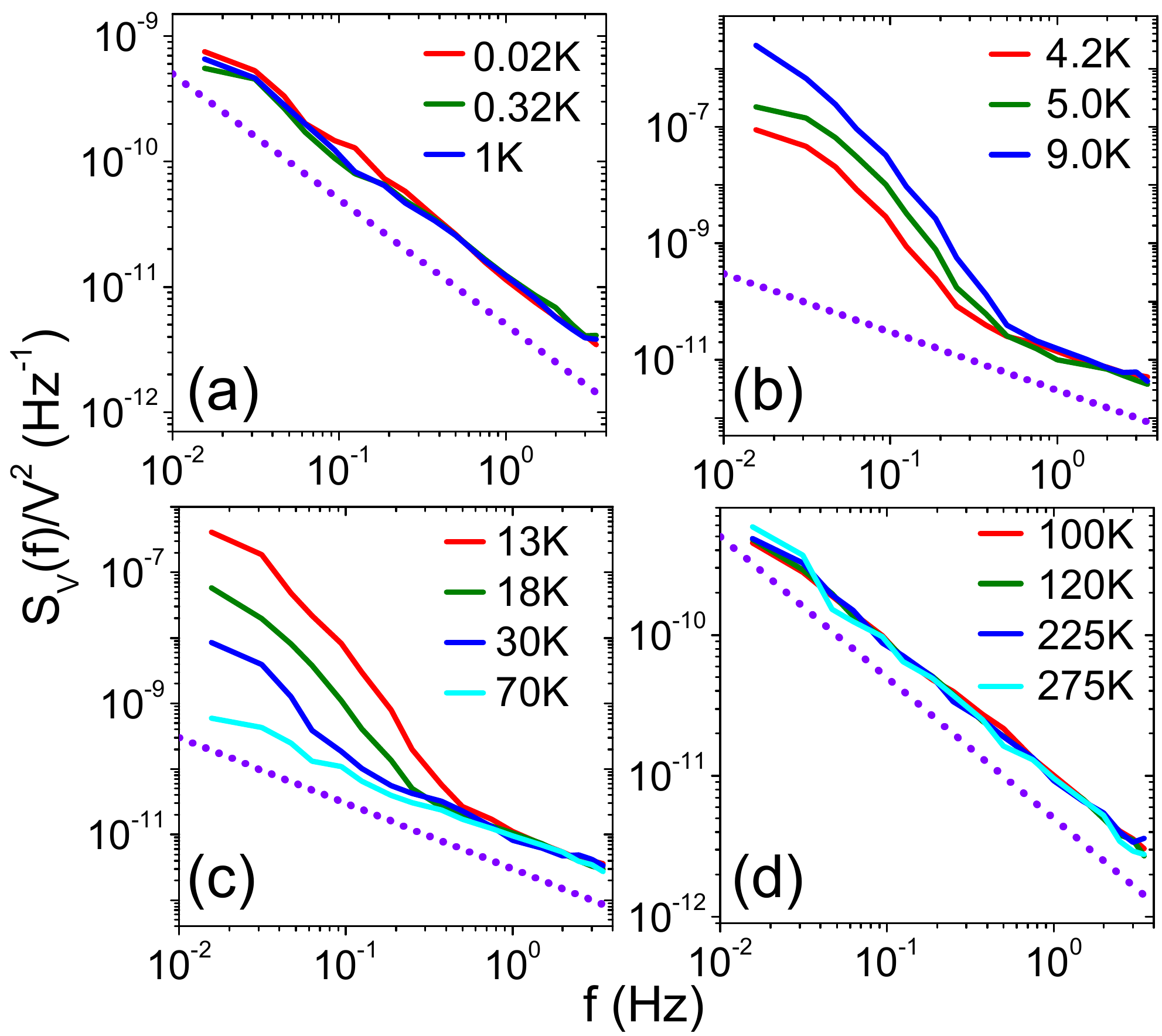}
		\small{\caption{ (color online) Normalized PSD of voltage fluctuations  $S_V(f)/V^2$ plotted as a function of frequency at a few representative temperatures in the four temperature regimes as described in the text. The dotted line in all four panels represents the $1/f$ noise. 
				\label{fig:spectrum}}}
	\end{center}
\end{figure}
\begin{figure*}[t!]
	\begin{center}
		\includegraphics[width=0.95\textwidth]{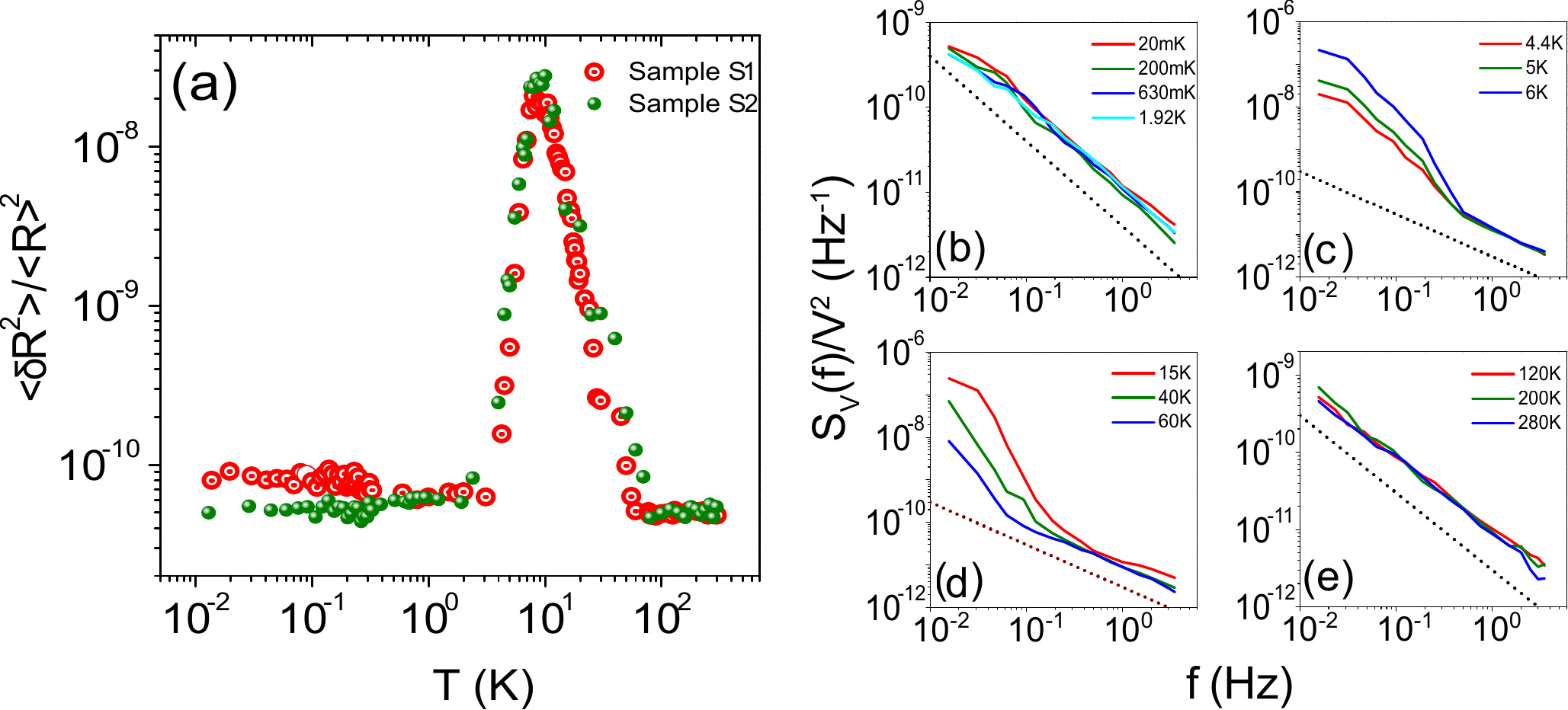}
		\small{\caption{ (color online) (a) Plot of \noise as a function of noise for the two samples - sample S1 (red circles) and sample S2 (olive filled circles). (b)-(e) Plots of \PSD~ of voltage fluctuations measured for sample S1 - the data are very similar to what was obtained for the sample S2 (see figure~\ref{fig:spectrum}).
				\label{fig:12PSD}}}
	\end{center}
\end{figure*} 

Figure~\ref{fig:noise} shows a plot of the relative variance of resistance fluctuations as a function of temperature and  figure~\ref{fig:spectrum} shows plots of PSD, $S_V(f)$ (scaled by $V^2$)  at a few representative temperatures for sample S2. The corresponding data for sample S1 are shown in Figure~\ref{fig:12PSD}. The data were collected over multiple thermal cycling of the devices from room temperature down to 10~mK to confirm reproducibility.   We found that the data obtained for both the samples were quantitatively very similar, hence we concentrate on the data from only sample S2 hereafter in this article. From the temperature dependence of the magnitude of \noise (Figure~\ref{fig:noise}), and the shape of the spectrum (Figure~\ref{fig:spectrum}), we conclude that there are four distinct temperature regimes, as indicated in figure~\ref{fig:noise}. Note that the temperatures regimes defined by us are quite similar to what had been found previously based on measurements of Hall co-efficient and thermopower~\cite{Sluchanko1999}.

At high temperatures (regime IV: $T$~>~80~K), the system behaves like a semi-metal~\cite{PhysRevB.3.2030, PhysRevLett.112.226402}. Over this temperature range, the noise is $1/f$ in nature and is very weakly dependent on temperature in accordance with previous results of noise in semi-metals~\cite{:/content/aip/journal/jap/55/5/10.1063/1.333224}. We do not discuss further the data over this temperature regime in this letter. In the following sections, we discuss, in detail the possible origins of noise in the different  temperature regimes at lower temperatures (10~mK~<~$T$~<~80~K). \\
 \begin{figure}[t!]
 	\begin{center}
 		\includegraphics[width=0.43\textwidth]{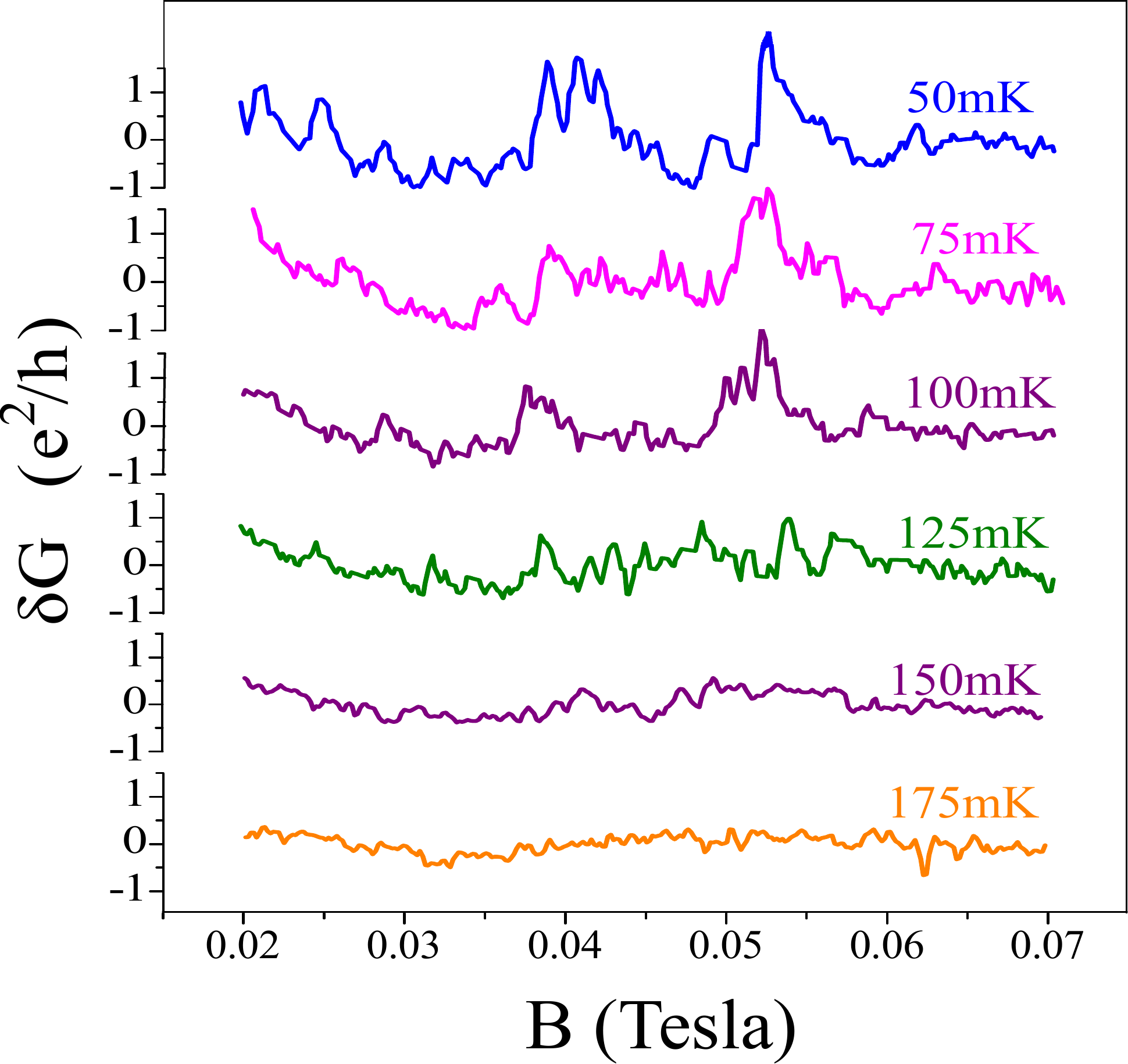}
 		\small{\caption{(color online) Plot of conductance fluctuations, $\delta$G as a function of magnetic field at a few different temperatures. The amplitude of UCF  was found to decrease with increasing temperature and individual UCF peaks could not be resolved above 150~mK. 
 				\label{fig:ucf1}}}
 	\end{center}
 \end{figure}
\begin{figure*}[t]
	\begin{center}
		\includegraphics[width=0.8\textwidth]{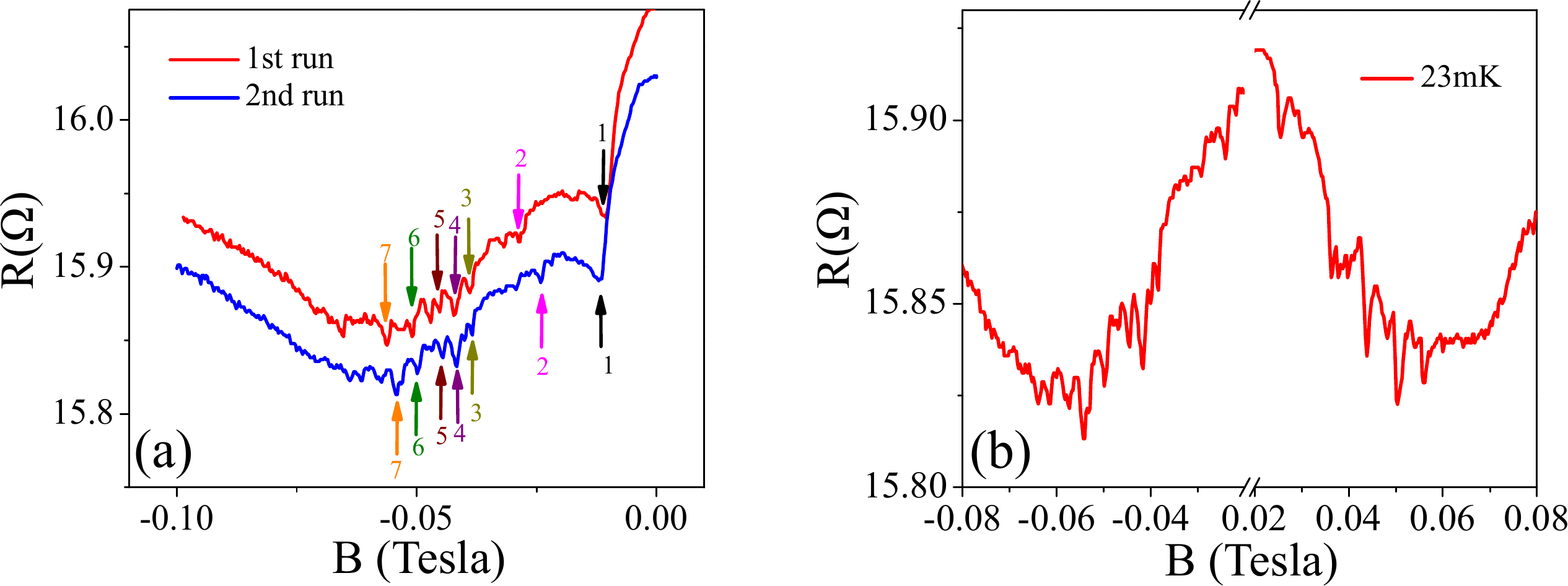}
		\small{\caption{ (color online) (a) Plot of magnetoresistance versus magnetic field for two consecutive runs taken at 23~mK. Peaks 1,3,4,5 and 6 appear at the same magnetic field values for both these set of data whereas peaks numbered as 2 and 7 have got shifted. (b) Plot of magnetoresistance versus magnetic field at 23~mK for positive and negative magnetic field sweeps.
				\label{fig:ucfinstability}}}
	\end{center}
\end{figure*} 
\section{Regime-I (Transport through only SS ($T$~<~3K))}: This is the regime of pure surface transport. Both local and non-local noise in this temperature regime have very weak temperature dependencies. The PSD, $S_V(f) \propto 1/f^\alpha$ with $\alpha\sim0.9-1.05$. 
Magnetoresistance shows signatures of weak anti-localization (WAL)~\cite{biswas2015robust,2013arXiv1307.4133T}. Observation of WAL implies strong effects of interference of electronic wave-functions on the transport properties. Another manifestation of quantum interference in mesoscopic systems at ultra-low temperatures is Universal Conductance Fluctuations (UCF)~\cite{PhysRevLett.55.1622}. UCF are sample specific aperiodic fluctuations in conductance observable by varying any parameter that affects the relative phase of the electronic wave-functions; \textit{e.g.} disorder configuration, magnetic field or chemical potential. Invoking the ergodic hypothesis, it can be argued that the same effect should be seen in the time trace of conductance fluctuations - at low temperatures, impurities or scattering centers in a sample can spontaneously rearrange themselves by quantum-mechanical tunneling giving rise to dynamic conductance fluctuations via UCF  with a $1/f$ spectrum~\cite{PhysRevB.56.15124}.

Low field magnetoconductance $\delta G(B)$ measured in regime-I shows UCF fingerprints~(figure~\ref{fig:ucf1}). The magnitude of UCF was found to decrease rapidly with increasing temperature, becoming indistinguishable from other sources of noise at temperatures above 150~mK. UCF should be symmetric with respect to the change in polarity of the magnetic field. Experimentally, we find that in SmB$_6$ this was not always the case. This was because of fact that the conductance of SmB$_6$ upon sweeping the magnetic field was seen to relax gradually over time scales of the order of minutes. This slow relaxation of the magnetoconductance can be due to a possible glassy state in SmB$_6$ arising due to the Ruderman-Kittel-Kasuya-Yosida (RKKY) interactions between local moments in the Kondo lattice~\cite{coleman2007heavy}. This glassy dynamic makes the overall conductivity a slowly varying function of time which makes it extremely difficult to find reproducible UCF fluctuations. Thus, even though we could find UCF peaks, the glassy background relaxations made the position of peak versus the magnetic field not quite as stable as in other materials. This is shown in figure~\ref{fig:ucfinstability}(a) where we present the magnetoresistance data from two consecutive magnetic field sweeps at 23~mK under identical measurement conditions. We find an overall similarity of fluctuations but the exact peak positions are not always reproducible. Some of the magnetoresistance peaks appear at exactly the same magnetic field values in both runs (e.g. those marked 1, 3, 4, 5 and 6) whereas a few other peaks got shifted. In figure~\ref{fig:ucfinstability}(b) we show the magnetoresistance data for both positive and negative field sweeps which are symmetric to the degree of instability as explained figure~\ref{fig:ucfinstability}(a). It should be noted that there are previous reports~\cite{wolgast2015magnetotransport,nakajima2016one} of hysteresis in magnetoresistance in SmB$_6$. Authors of reference~\cite{wolgast2015magnetotransport} suggest that magnetocaloric effect or magnetic impurity scattering due to the presence of samarium oxide (Sm$_2$O$_3$) layer are possible causes of the hysteresis in magnetoresistance. On the other hand, the authors of reference~\cite{nakajima2016one} propose that the hysteresis is due to the appearance of a magnetic ordered state at low temperatures. More theoretical and experimental work is required to understand the effect of this possible magnetic ordering on the glassy state observed by us.

For a 2-dimensional system of length $l$, the rms  conductance fluctuations in the symplectic symmetry class (which is the relevant symmetry class for surface states of 3-dimensional topological insulators)is given by ~\cite{1367-2630-14-10-103027}:
\begin{equation}
\langle(\delta G)^2\rangle^{1/2} =\sqrt{\frac{3}{\pi}} \frac{e^2}{h}\frac{l_\phi}{l}.
\label{ucf}
\end{equation}
Note that Eqn.~\ref{ucf} is valid in the limit $l \gg l_\phi$ which is the case for our samples. 
The value of  $l_\phi$ at 20~mK obtained from the low field magnetoconductance fluctuation data using Eqn.~\ref{ucf} is $\sim$1400~nm which is quite close to the value found from WAL ($\sim$1200~nm)~\cite{biswas2015robust,2013arXiv1307.4133T} showing that these are indeed UCF. Also, the value of  $\langle(\delta G)^2\rangle^{1/2}$ obtained at 20~mK from the time series of resistance fluctuations [using the data presented in Fig.~\ref{fig:noise}] is 0.0012~$e^2/h$ which matches quite well with the variance $\langle(\delta G)^2\rangle^{1/2}$ obtained from UCF measurements (0.0014$e^2/h$).  These observations strongly suggest that the noise measured over this temperature regime originates predominantly from the quantum interference of electronic wave functions and that the transport is confined to a 2-dimensional region.

\begin{figure}[t!]
	\begin{center}
		\includegraphics[width=0.45\textwidth]{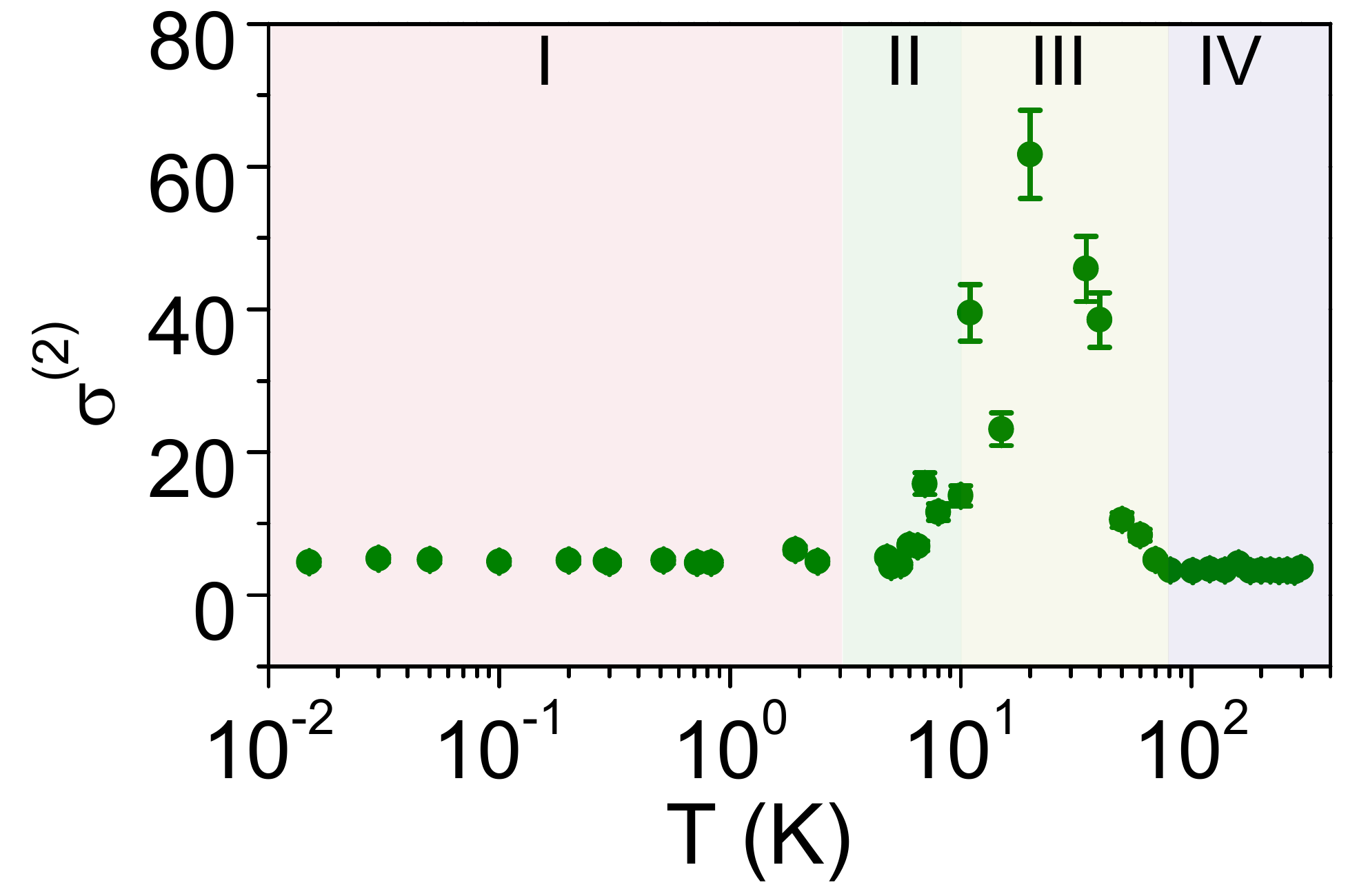}
		\small{\caption{(color online) Plot of the normalized second spectrum $\sigma^{(2)}$ as a function of temperature. Error bars were calculated as standard deviations from measurements of $\sigma^{(2)}$ over 50 time windows.  The shading indicates the four different temperature regimes as explained in detail in the text. 
				\label{fig:2spec}}}
	\end{center}
\end{figure}

\begin{figure*}[t!]
	\begin{center}
		\includegraphics[width=0.75\textwidth]{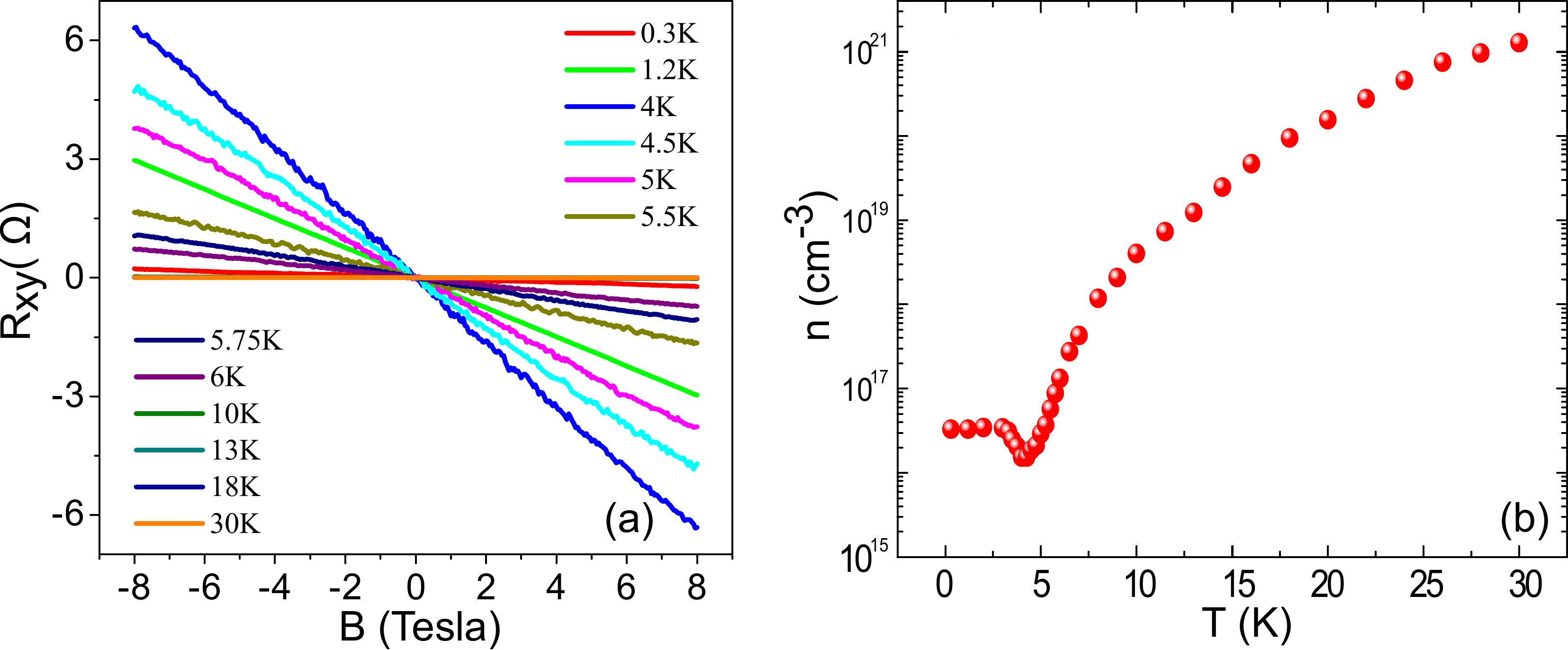}
		\small{\caption{ (color online) (a) Plot of transverse resistance $R_{xy}$ as a function of perpendicular magnetic field measured over the temperature range 0.3~K $\leqslant$ $T$ $\leqslant$ 30~K.  (b) charge carrier number density as a function of temperature - the data have been calculated from the measured $R_{xy}$ assuming a 3-dimensional transport. Note that this description breaks down below 3~K - the number density over this temperature regime corresponds to a surface number density $\sim 1.6\times 10^{15}$ cm$^{-2}$.
				\label{fig:Hall}}}
	\end{center}
\end{figure*} 

\section{Regime-III (Bulk transport in Kondo Hybridization regime ($10~K~<T<~80~K$))}: In this temperature regime we observe that the PSD deviates from $1/f$ dependence at frequencies below $\sim$~0.5~Hz (figure~\ref{fig:spectrum}(c)). The noise increases rapidly by about three orders of magnitude with decreasing temperature attaining a maximum value of  $\sim 2\times 10^{-8}$ at 10~K. This is qualitatively similar to previous observation~\cite{glushkov2003low}. The large value of $dR/dT$ in this regime naturally raises the question whether the measured resistance fluctuations can arise from temperature fluctuations. According to the thermal fluctuation model of Voss and Clarke~\cite{voss1976flicker}, temperature fluctuations can give rise to low frequency $1/f$ noise with \PSD given by 
	\begin{eqnarray}
	S_{V}(f)=V^{2} {\beta}^{2}<(\Delta T)^{2}>
	\end{eqnarray}  
where $\beta=1/R(dR/dT)$. Using $<(\Delta T)^{2}>=k_{B}T^{2}/C_{V}$ and getting $\beta=1/R(dR/dT)$ from the resistance vs temperature data, we can estimate $S_{V}(f)/V^{2}$.  The measured noise in our sample is maximum at 10~K. At this temperature the reported value of $C_{V}$ is $ \sim 0.05$ Jmol$^{-1}$K$^{-1}$~\cite{phelan2014correlation}. Using this value we get $S_{V}(f)/V^{2}=4 \times 10^{-17}$~V$^2$/Hz which is orders of magnitude smaller than our experimentally measured value of $2.5 \times 10^{-6}$~V$^2$/Hz. Hence, we can rule out thermal fluctuations as the primary cause of the large noise seen in SmB$_6$ over this temperature range. Another possible origin of resistance fluctuations is defect motion. Pelz and Clarke~\cite{pelz1987quantitative} showed that defect motion can give rise to low frequency noise with a $1/f$ \PSD. Their model, called the Local Interference (LI) model predicts a noise magnitude:
	\begin{eqnarray}
	\dfrac{\langle\delta R^{2}\rangle}{\langle R\rangle^{2}} = \dfrac{1}{N} (n_{a}l_{mfp} \zeta \chi)^{2}\dfrac{n_{m}}{n_{a}};
	\label{eqn:LI}
	\end{eqnarray}
where $N$ is the number of atoms in the sample,    $n_{a}$= $\dfrac{N}{V}$, $V$ is the sample volume, $l_{mfp}$ is the mean free path of charge carriers, $\zeta$ is the anisotropy parameter, $\chi$ is the average defect cross-section and $n_{m}$ is the number density of mobile defects. The estimated value of  $l_{mfp}$  in our sample is $\sim$~28~nm. $\zeta$ generally takes value $\sim 0.1 -0.2$. Defect cross-section could also be calculated using $\chi \sim 4\pi/k_{F}^{2}$. Simple free electron model gives $k_{F}=(3\pi^{2}n)^{1/3}$ which gives us $\chi \sim 5\times 10^{-17}$ at 10~K. Using these values in Eqn.~\ref{eqn:LI}, we estimate that we must have $\dfrac{n_{m}}{n_{a}} \sim 415$ to account for the measured noise. This is a physically impossible result since $n_{m}$ can not exceed $n_{a}$. This leads us to believe that the LI defect fluctuation model can not explain the magnitude of noise measured in the high-temperature regime. Below we propose a likely scenario that can explain the large non-$1/f$ noise observed in this temperature regime.
    
As the temperature is lowered below $\sim90$~K, the  Kondo hybridization gap begins to develop~\cite{zhang2013hybridization,xu2014exotic} making the bulk of the crystal insulating.  The number of itinerant electrons taking part in screening the local moments increases resulting in an exponential decrease in the carrier number density. These screened local moments act as strong scattering centers for the remaining itinerant electrons giving rise to the sharp resistivity increase. In this state, quantum fluctuations of the magnetic and electronic degrees are strongly coupled causing the screened local moments to undergo slow fluctuations. It has been predicted that RKKY interaction between these fluctuating local moments via the intervening screening electron cloud can result in a spin-glassy state~\cite{coleman2007heavy,binder1986spin}. We propose that it is this glassy-dynamics which is responsible for the large increase in low-frequency noise in this temperature regime. 
 
 There are two distinct measurable signatures of slow glassy dynamics in electrical noise: (i) the PSD would deviate significantly from $1/f$ form at lower frequencies (see figure~\ref{fig:spectrum}(c)), and (ii) the PSD should show temporal fluctuations~\cite{WEISSMAN199287}. This, in turn, would introduce significant non-Gaussian components (NGC) in the resistance fluctuations which can be experimentally probed by studying the higher-order statistics of the measured noise. To test whether the noise in this temperature range originates from glassy dynamics we calculated the fourth moment of voltage fluctuations - what is commonly known in as `Second spectrum'~\cite{koushik2013correlated,  restle1985non, seidler1996dynamical}. Physically, second spectrum represents the fluctuations in the PSD with time in the chosen frequency octave. Operationally, the second spectrum is the Fourier transform of the four point voltage-voltage correlation function calculated over a chosen frequency octave ($f_l$, $f_h$). It can be defined as:
\begin{equation}
S_V^{f_1}(f_2)=\int_0^\infty \langle\delta V^2(t)\rangle\langle\delta V^2(t+\tau)\rangle cos(2\pi f_2\tau)d\tau
\end{equation}
where $f_1$ is the center frequency of a chosen octave and $f_2$ is the spectral frequency. Details of the calculation method are given in ref.~\citep{koushik2013correlated}. The  normalized value of the second spectrum $\sigma^{(2)}$ is given by
\begin{equation}
\sigma^{(2)}=\int_0^{f_h-f_l}S_V^{f_1}(f_2)df_2/[\int_{f_l}^{f_h}S_V(f)df]^2
\end{equation}
For Gaussian processes, the value of  $\sigma^{(2)}$ is identically three. In figure~\ref{fig:2spec}  we plot the value of $\sigma^{(2)}$ calculated over the octave 47~mHz - 94~mHz.  The sharp deviation from the Gaussian value in the temperature regime-III is indicative of long-range correlations in the system suggesting the presence of glassy states. Having said that, we believe that more work is necessary to unambiguously pin down the source of the large non-$1/f$ noise seen over this temperature regime.

Glassy dynamics couples to the voltage fluctuations in a system through fluctuations in the carrier mobility. In the case of noise originating from mobility fluctuations, the relative variance of resistance fluctuations scales inversely as the carrier density, \noise $\propto 1/n$. To understand if this is the source of increased noise in the system, we performed Hall measurements in the local configuration. Typical plots of $R_{xy}$ are shown in figure~\ref{fig:Hall}(a). Figure~\ref{fig:Hall}(b) shows the charge carrier density $n$ extracted from the measured  $R_{xy}$ assuming a 3-dimensional transport. The low temperature (<~3~K) saturation of number density corresponds to surface states with a carrier density of $\sim 1.6\times 10^{15}$ cm$^{-2}$. Charge carrier density in the bulk, on the other hand, has a thermally activated behavior with a temperature dependent activation energy~\cite{zhang2013hybridization, allen1979large} . Figure~\ref{fig:noise}  shows the fit to the measured \noise using this form (pink solid line)- the excellent match supports the idea that noise in this temperature regime arises from mobility fluctuations of the charge carriers. 

Note that, although ARPES measurements find the presence of surface states till about 40~K~\cite{xu2014exotic}, transport measurements did not show any indication of SS above 10~K. No non-local signal could be detected over this temperature range, either in voltage or in voltage fluctuations confirming that transport occurred only through the bulk.

\begin{figure*}[t!]
	\begin{center}
		\includegraphics[width=0.95\textwidth]{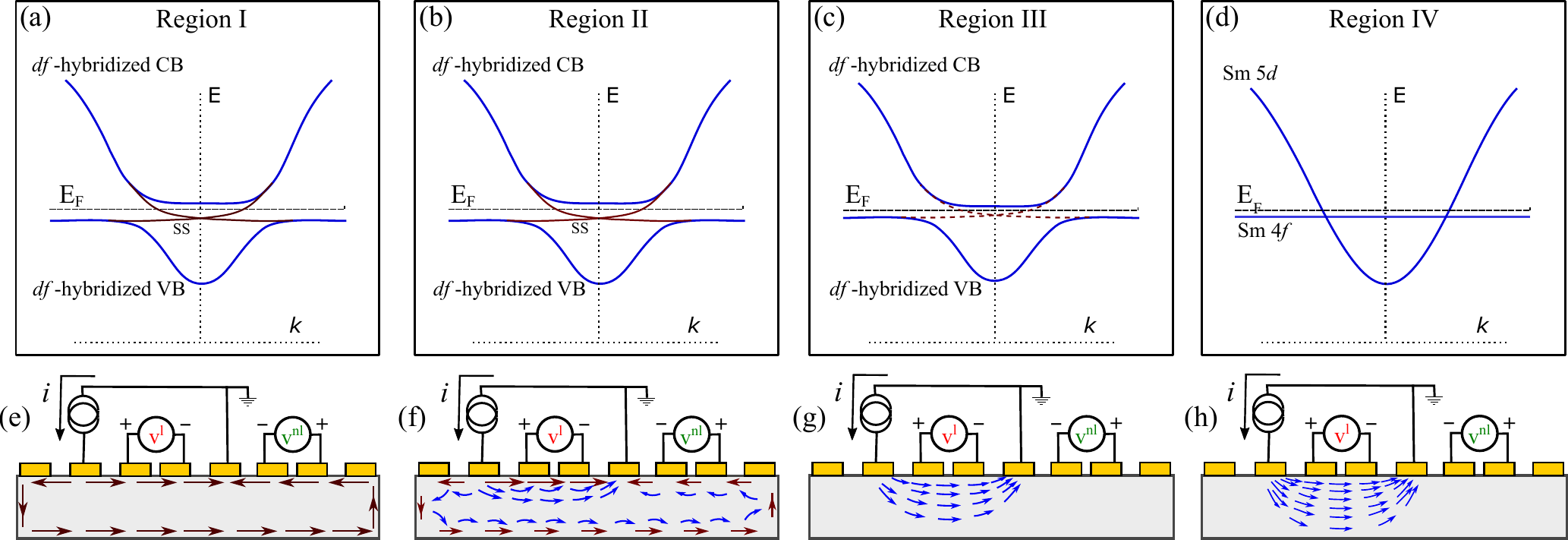}
		\small{\caption{ (color online) (a)-(d) Sketch of the possible band structure over the different temperature regimes. (e)-(h) Device schematic showing the current propagation modes in the different temperature regimes. In regime I, the entire current flows through the SS. In regime II, the bulk resistance is comparable to that of the SS and current flows both the through the bulk of the crystal and through the SS. In regime III and IV there is no evidence of the presence of SS  from transport measurements and the entire current flows through the bulk of the sample.  
				\label{fig:schematic}}}
	\end{center}
\end{figure*}
\section{Regime-II (Bulk and surface combined transport (3~K~<~$T$~<~10~K))}:  Over this temperature range, the value of $R_b/R_s$ decreases rapidly with increasing temperature. Consequently, the ratio of the current flowing through the bulk to that flowing through the SS increase rapidly. This is the regime of mixed surface and bulk transport where the current flows through both the bulk and SS [see Fig.~\ref{fig:RT}(b)].  
The noise measured in both local and non-local configurations yield very similar values. This can be understood by noting that if the resistance fluctuations in the surface states and in the bulk regions are uncorrelated, the total noise in this regime can be evaluated using a simple parallel channel model:
\begin{eqnarray}
\dfrac{\langle\delta R^{2}\rangle}{\langle R\rangle^{2}}=\left( \frac{R}{R_b}\right)^2 \dfrac{\langle\delta R_s^{2}\rangle}{\langle R_s\rangle^{2}} + \left(\dfrac{R}{R_s}\right)^2 \dfrac{\langle\delta R_b^{2}\rangle}{\langle R_b\rangle^{2}} 
\label{eqn:noisediv}
\end{eqnarray}
In figure~\ref{fig:noise} we show a plot of  \noise estimated using this formalism (blue solid line). The quantities  $\langle\delta R_s^2\rangle/ \langle R_s\rangle^2$ and $\langle\delta R_b^2\rangle/ \langle R_b\rangle^2$  used in equation~\ref{eqn:noisediv} were estimated by using the temperature dependencies of these parameters measured in regime-I and regime-III respectively. The excellent match between the estimated noise with both the measured local and non-local noise confirms the accuracy of the parallel transport channel model.  At temperatures higher than 10~K, the resistance of the bulk becomes negligible compared to that of the surface states and the entire current flows only through the bulk. The system then enters regime-III which is the domain of pure bulk transport.

To conclude, we have studied the charge-carrier scattering dynamics in high-quality single crystal samples of SmB$_6$ starting from room temperature down to ultra-low temperatures and have probed the evolution of the system from semi-metallic state to Kondo correlated state and finally to metallic surface state. In figure~\ref{fig:schematic} we show schematically the possible evolution of the band structure in SmB$_6$ with decreasing temperature. We also show the deduced current flow patterns for every temperature regime. In regime I, $R_b$ is very large as compared to $R_s$. Consequently, the entire current flows through SS.  At low temperatures, the transport in the system takes place through surface states and the measured noise in this range arises due to universal conductance fluctuations. Our observation of signatures of UCF, which are a manifestation of quantum interference of electron wave function in a two-dimensional system, is another independent proof of the existence of metallic surface states in this material.  With increasing temperature (regime II), the bulk resistance becomes comparable to that of the SS and current flows both through the bulk of the crystal and through the surface channels. In regime III, ARPES measurements indicate the presence of surface states. The bulk resistance over this temperature range is much smaller compared to that of the surface states ($R_b \ll R_s$). Consequently, the entire current flows through the bulk of the sample and we do not get any signature of surface transport.  At temperatures higher than the Kondo temperature, electrical transport is through the bulk and the noise measured in this regime arises due to mobility fluctuations. We find signatures of glassy dynamics in this temperature range. The noise measured in this regime arise due to mobility fluctuations and reveals signatures of glassy dynamics. At very high temperatures (regime IV, $T>T_K$), only the electrons in the dispersive 5d orbital contribute to electrical transport and the entire current flows through the bulk of the sample.  Unlike the case of topological insulators based on bismuth dichalcogenides~\cite{bhattacharyya2015bulk}, in SmB$_6$ we find that the noise in SS and bulk are uncorrelated - at ultra-low temperatures bulk has no discernible contribution to electrical transport making SmB$_6$ ideal for probing the physics of topological surface states.

\begin{acknowledgments}
A.B. acknowledges funding from Nanomission, Department of Science \& Technology (DST) and Indian Institute of Science. The work at the University of Warwick was supported by a grant from EPSRC, UK, Grant EP/L014963/1.
\end{acknowledgments}

\renewcommand{\thefigure}{A\arabic{figure}}
\setcounter{figure}{0}
\begin{figure*}[t!]
	\begin{center}
		\includegraphics[width=0.5\textwidth]{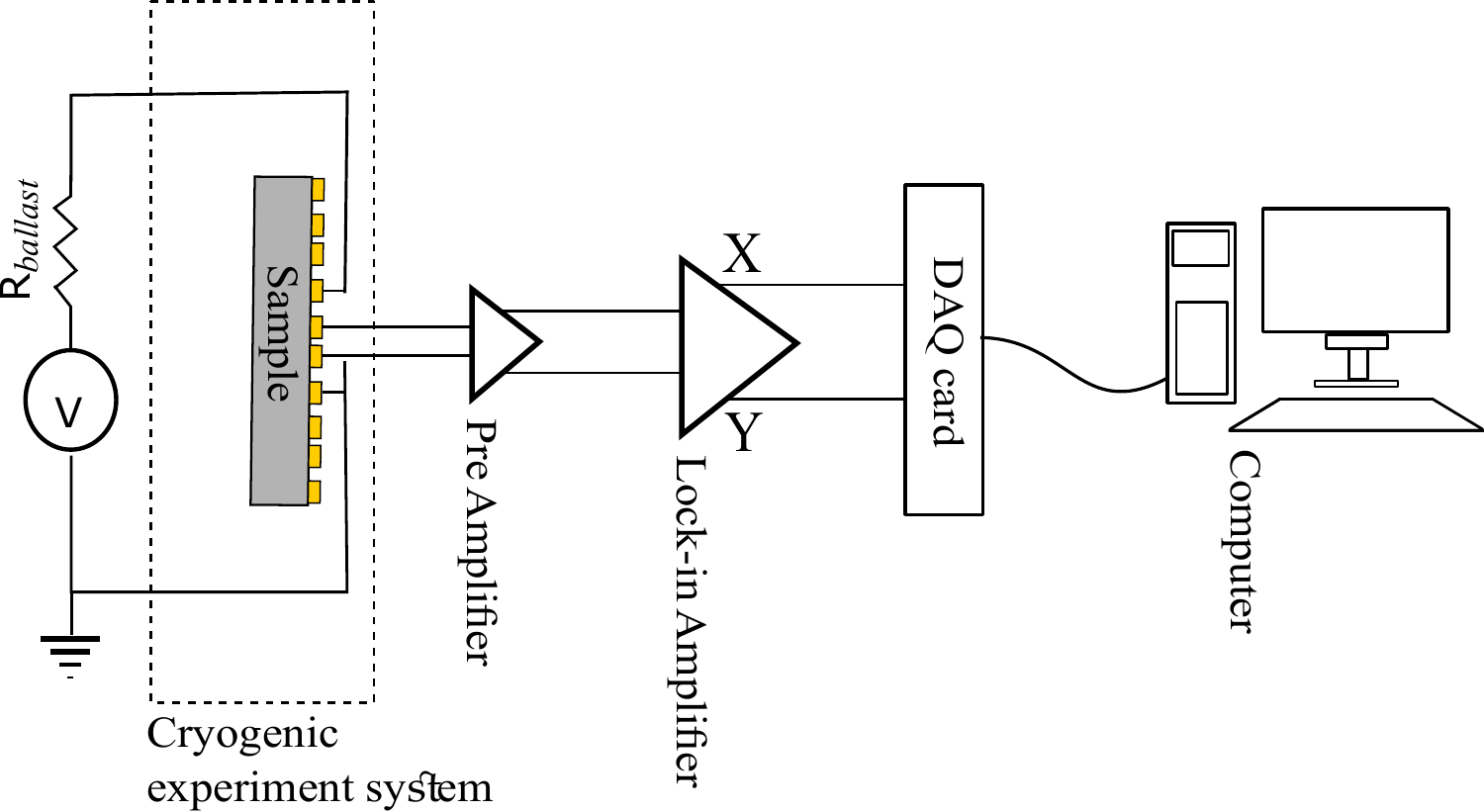}
		\small{\caption{ (color online) Schematic of setup used for low frequency noise measurements.  
				\label{fig:setup}}}
	\end{center}
\end{figure*} 
\begin{figure*}[t!]
	\begin{center}
		\includegraphics[width=0.85\textwidth]{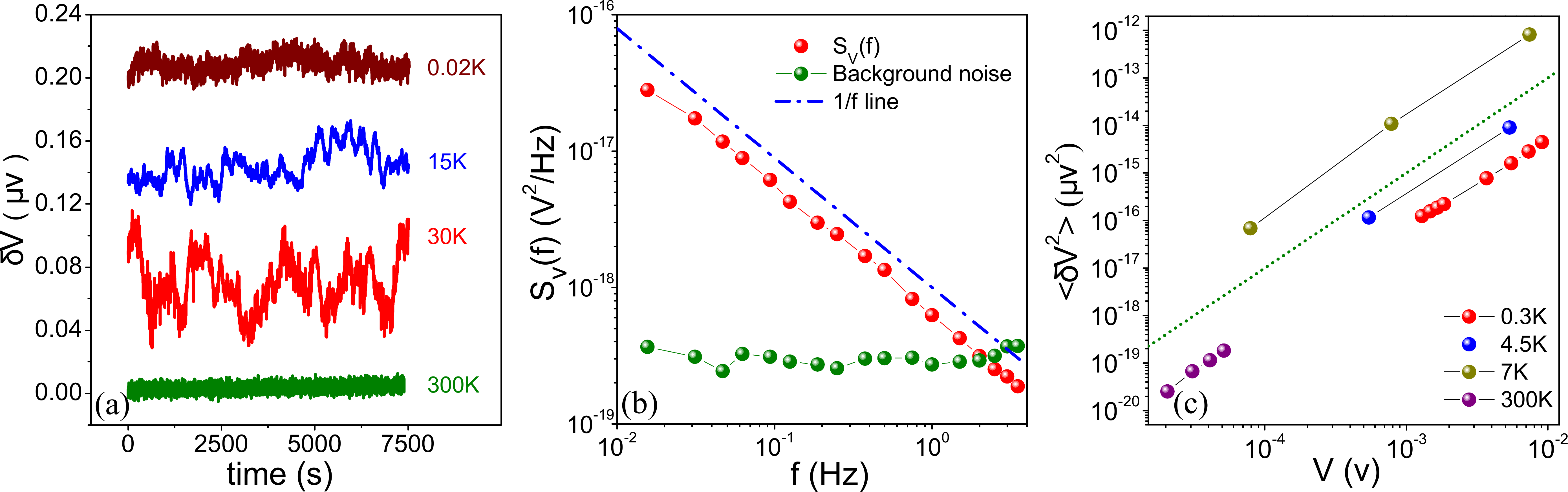}
		\small{\caption{ (color online) (a) Time series of local voltage fluctuations, $\delta V^l$  at a few representative temperatures - 300~K (olive line), 30~K (red line), 15~K (blue line) and 0.02~K (maroon line). The data have been vertically offset for clarity. (b) Plot of the power spectral density of voltage fluctuations, $S_{V}(f)$ as a function of frequency. The red curve is the $1/f$ noise while the green curve shows the \PSD of white background noise. The blue dashed line shows $\propto 1/f$ function. (c) Plot of the variance of voltage fluctuations ($<\delta V^{2}>$) as a function of $V$ in log-log scale. The slopes of all the plots are very close to 2 showing that $<\delta V^{2}>~ \propto V^2$. The olive dashed line indicates the $\propto V^{2}$ function. 
				\label{fig:PSDs}}}
	\end{center}
\end{figure*} 

\section* {appendix: Details of noise measurement technique}

Figure ~\ref{fig:setup} shows a schematic of the setup used to measure low-frequency voltage noise. An SR830 dual-channel lock-in-amplifier (LIA) was used to bias the sample at a carrier frequency $f_{0}$ $\sim$ $228~Hz$. A large ballast resistor R$_{ballast}$ of few orders of magnitude higher value than the sample resistance was connected in series with the sample to ensure a constant current flow. Standard four probe geometry was used to measure the voltage difference between two probes which was coupled to the input of the LIA using a low-noise voltage pre-amplifier. The input signal to the LIA was digitally off-set to obtain the fluctuation $\delta V$ of the voltage about the average value $<V>$.  The dc output of the LIA was recorded by a fast 16 bit analog to digital conversion card at a sampling rate of 2048 points per second.  Total time duration of each of our time series data was 32 minutes. Figure ~\ref{fig:PSDs}(a) shows typical time traces of voltage fluctuations measured at a few representative temperatures.
The recorded time series was digitally anti-alias filtered and the power spectral density of voltage fluctuations obtained from it using the Welch's method of averaged periodogram. The spectral decomposition of the measured time series was done over the frequency window 0.01563~Hz\ $\leqslant$ $f$ $\leqslant$~3.5~Hz. Subtracting the \PSD~ of the quadrature component of the LIA output from the in-phase component gives the \PSD~   $S_{V}(f)$ of the voltage fluctuations coming from the sample  (figure ~\ref{fig:PSDs}(b)). It was verified that at every temperature $S_V(f) \propto V^2$ in the operating voltage excitement range thus ensuring linear transport regime (figure ~\ref{fig:PSDs}(c)). The relative variance ( \noise ) was calculated using:

\begin{equation}
\dfrac{\langle\delta R^{2}\rangle}{\langle R\rangle^{2}}=  \dfrac{1}{V^{2}} \int\limits_{0.01563Hz}^{3.5Hz} S_{V}(f) df.
\end{equation}

%

\end{document}